\documentstyle[aps,twocolumn,epsfig]{revtex}
\draft
\newcommand{\nn}{\nonumber}
\newcommand{\be}{\begin{equation}}
\newcommand{\ee}{\end{equation}}
\newcommand{\bea}{\begin{eqnarray}}
\newcommand{\eea}{\end{eqnarray}}
\newcommand{\h}{Hamiltonian }
\renewcommand{\o}{\omega}
\newcommand{\hb}{\hbar}
\newcommand{\la}{\langle}
\newcommand{\ra}{\rangle}
\newcommand{\half}{\frac{1}{2}\,}
\renewcommand{\inf}{\infty}
\newcommand{\etal}{{\em{et}}\,{\em{al.}}}
\newcommand{\nb}{\bar{n}}

\begin{document}

\title{Eigenstates of a Small Josephson Junction Coupled to a 
Resonant Cavity}
\author{W.~A.~Al-Saidi\cite{email1} and D.~ Stroud\cite{email2}}
\address{
Department of Physics,
The Ohio State University, Columbus, Ohio 43210}
\date{\today}
\maketitle
\begin{abstract}
We carry out a quantum-mechanical analysis of a small Josephson
junction coupled to a single-mode resonant cavity. We find that the
eigenstates of the combined junction-cavity system are strongly
entangled only when the gate voltage applied at one of the
superconducting islands is tuned to certain special values.  One such
value corresponds to the resonant absorption of a single photon by
Cooper pairs in the junction. Another special value corresponds to a
{\em two-photon} absorption process.  Near the single-photon resonant
absorption, the system is accurately described by a simplified model
in which only the lowest two levels of the Josephson junction are
retained in the Hamiltonian matrix.  We noticed that this
approximation does not work very well as the number of photons in the
resonator increases.  Our system shows also the phenomenon of
``collapse and revival'' under suitable initial conditions, and our
full numerical solution agrees with the two level approximation
result.
\end{abstract}

\pacs{PACS numbers:  03.65.Ud, 03.67.Lx, 42.50.-p, 74.50.+r}

\section{Introduction}

Circuits involving small Josephson junctions have the potential to
behave as macroscopic two-level systems which can be externally
controlled. Indeed, several recent experiments have demonstrated that
such a system can be placed in a coherent superposition of two
macroscopic quantum states.  The experiments have involved small
superconducting loops, and also so-called Cooper pair boxes.  In the
former case, it was shown experimentally\cite{friedman,vanderwal} that
the loop could exist in a coherent superposition of two macroscopic
states of different flux through the loop.  In the case of the Cooper
pair box, the system was placed into a superposition of two states
having different numbers of Cooper pairs on one of the superconducting
islands\cite{bouchiat,nakamura1,nakamura2,zorin}.  Such a coherent
superposition had been proposed by several
authors\cite{mooij,feigelman}.

One possible application of such two-state systems is as a quantum bit
(qubit) for use in quantum computing\cite{steane}.  At sufficiently
low temperatures, this Josephson qubit will have little dissipation,
and hence may be coherent for a reasonable length of time, as is
required of a qubit.  But in order for it to be potentially useful in
computation, it must be possible to prepare entangled states of two
such qubits - that is, states of the two qubits which cannot be
expressed simply as product states of the two individual qubits.

Such entangled states have now been created and manipulated in systems
involving atoms in a high-Q cavity\cite{haroche}. Only recently has
there been any theoretical study of entangled states involving
Josephson junction circuits. Shnirman {\etal} \cite{shnirman} have
considered low-capacitance Josephson junctions coupled to an
LC-circuit in the two level approximation. Buisson and
Hekking\cite{hekking} have studied the states of a single Cooper pair
box coupled to the photon mode in an electromagnetic resonant
cavity. Both \cite{shnirman}, and \cite{hekking} considered junctions
such that the charging energy is much larger than the Josephson
energy. Everitt \etal \cite{everitt} have carried out a similar
investigation of the states of a small superconducting quantum
interference device (a SQUID - i. e., a small superconducting loop)
coupled to a resonant cavity. In all of these studies, entangled
states of the photon and the Josephson junction emerged naturally from
the quantum-mechanical analysis.

In the present paper, we carry out a full quantum-mechanical
investigation of a model for a Josephson junction coupled to a
single-mode resonant photonic cavity.  Our model Hamiltonian is, as we
demonstrate here, equivalent to that studied in Ref.\ \cite{hekking},
via a canonical transformation.  However, we consider a different
parameter region where the charging and Josephson energy of the
junction are comparable in magnitude.  Also we go beyond the analysis
of that paper by including a large number of eigenstates of both the
Josephson and the photon Hamiltonian, in calculating the energy levels
of the coupled system.  When the system is at resonance, we find that
the entangled states are indeed, as expected, well described by a
basis of two states of the combined system.  We also find a variety of
entangled states, depending on the value of the control parameter (an
offset voltage), including, at certain values, an eigenstate which is
a superposition of two states in which the number of photons differs
by {\em two}.  We have also verified that our system shows the
phenomenon of ``collapse and revival'' similar to that seen in quantum
optics\cite{eberly}.  Our approach is similar to that applied by
Everitt \etal \cite{everitt} to a SQUID ring in a resonant cavity,
though our Hamiltonian is not equivalent.

The remainder of this paper is organized as follows.  In Section II,
we give the model Hamiltonian, present several of its properties, and
describe our method of solving it.  The following section gives
numerical results obtained from the model.  The last section contains
a concluding discussion, an estimate for some of the parameters used in our
model, and a description of possible problems for future work.

\section{Model}

\subsection{Model Hamiltonian}

We consider an underdamped Josephson junction in a large-$Q$ electromagnetic
cavity which can support a single photon mode.  
The system is assumed to be described by 
the following model Hamiltonian:
\begin{equation}
{\mathcal{H}} = {\mathcal{H}}_{\mathrm{photon}} + {\mathcal{H}}_J + 
{\mathcal{H}}_C.
\end{equation}
Here ${\mathcal{H}}_{\mathrm{photon}}$ is the \h  of the cavity mode, 
which we express in
the form
$
{\mathcal{H}}_{\mathrm{photon}} = \hbar\,\omega\left(a^\dag a + 
{1}/{2}\right),
$
where $a^\dag$ and $a$ are the usual photon creation and annihilation
operators.  ${\mathcal{H}}_J$ is the Josephson coupling energy, which we write in
the form
$
{\mathcal{H}}_J = -J\cos\gamma,
$
where $J$ is the Josephson energy of the junction and $\gamma$ is the
gauge-invariant phase across the junction (defined more precisely below).
$J$ is related to $I_c$, the critical current of the junction, by 
$J = \hbar I_c/(2\,|e|)$.
Finally, ${\mathcal{H}}_C$ is the capacitive energy of the junction, which we write as
$
{\mathcal{H}}_C = \frac{1}{2}\,U\,(n - \bar{n})^2,
$
where $U=4\,e^2/C$ is the charging energy of the junction with capacitance
$C$; $n$ is an operator which represents the difference 
between the number of Cooper pairs on the two
superconducting islands which make up the junction; and
$\bar{n}$ is a tunable experimental parameter related
to the gate voltage which is applied at  one of the superconducting islands.

The gauge-invariant phase difference $\gamma$ is the term which couples the
Josephson junction to the cavity.  It may be written as
$
\gamma = \phi - ({2\pi}/{\Phi_0})\int {\bf A}\cdot d{\bf \ell} \equiv
\phi - A,
$
where $\Phi_0=hc/(2 e)$ is the flux quantum, $\phi$ is the phase difference across the junction in a particular gauge,
${\bf A}$ is the vector potential, and the line integral is taken across
the junction.  We assume that ${\bf A}$ arises
from the electromagnetic field of the cavity normal mode.  In Gaussian
units, and assuming the Coulomb gauge (${\bf \nabla}\cdot {\bf A} = 0$),
this vector potential takes the form
$
{\bf A} = \sqrt{{h\,c^2}/{\omega\, V}}\left( a + a^\dag\right)
{ \bf \hat{\epsilon}},
$
where ${\hat{\epsilon}}$ is the unit polarization vector of the cavity
mode, and $V$ is the volume of the cavity. Here we assumed that the
junction dimensions are much smaller
than the wavelength of the resonant cavity mode, so that the cavity electric
field is approximately uniform within the junction.
Given this representation for ${\bf A}$, the phase factor $A$ can be written
as
$
A = ({g}/{\sqrt{2}})(a + a^\dag),
$
where 
\be
g = \frac{4 \, \sqrt{\pi} |e|}{ \sqrt{ \hb\,{\omega}\,V}} \,\ell_{\|}, \label{eq:g}
\ee
and $\ell_{\|}= {\bf \hat{\epsilon}}\cdot {\vec \ell}$ is the
effective length across the junction parallel to $\hat{{\bf
\epsilon}}$.  The definition of the Hamiltonian is completed by
writing down the commutation relations for the various operators.  The
only nonzero commutators are $[a, a^\dag] = 1,$ and $ [n, \phi] = -i.
$
 
It is useful to make a change of variables 
$
p = i\sqrt{\hbar\,\omega/{2}}\left(a^\dag - a\right); 
q = \sqrt{{\hbar}/{(2\,\omega)}}\left(a^\dag + a\right).
$
Then $p$ and $q$ satisfy the canonical commutation relation
$[p, q] = -i\,\hbar,$ 
and ${\mathcal{H}}_{\mathrm{photon}}$ can be expressed in a form which makes clear that it
represents the Hamiltonian of a harmonic oscillator:
\begin{equation}
{\mathcal{H}}_{\mathrm{photon}} = \frac{1}{2}\left(p^2 + \omega^2 q^2\right).
\end{equation}
In terms of these new variables, the gauge-invariant phase difference
is given by
$\gamma = \phi - g\,\sqrt{{\omega}/{\hbar}}\,q.$

The Hamiltonian ${\mathcal{H}}$ can now be expressed conveniently as the sum of
three parts: 
\begin{equation}
{\mathcal{H}} = {\mathcal{H}}_{\mathrm{JJ}} + {\mathcal{H}}_{\mathrm{photon}} + {\mathcal{H}}_{\mathrm{int}},
\label{eq:htot}
\end{equation}
where 
\begin{equation}
{\mathcal{H}}_{\mathrm{JJ}} = \half U\,(n - \bar{n})^2 -J\cos\phi
\end{equation}
and 
\begin{equation}
{\mathcal{H}}_{\mathrm{int}} = -J\,(\cos\gamma - \cos\phi).
\label{eq:hint}
\end{equation}
In the absence of ${\mathcal{H}}_{\mathrm{int}}$ the Hamiltonian is 
the sum of two
independent parts, corresponding to the Josephson junction and the
resonant photon mode.  ${\mathcal{H}}_{\mathrm{int}}$ couples these 
two together.

\subsection{Method of Solution}

It is convenient to diagonalize
${\mathcal{H}}$ in a complete basis consisting of  the direct product of the 
eigenfunctions of ${\mathcal{H}}_{\mathrm{JJ}}$ and 
${\mathcal{H}}_{\mathrm{photon}}$.

The eigenfunctions of ${\mathcal{H}}_{\mathrm{photon}}$ are, of course, 
harmonic oscillator eigenstates; the n$^{th}$ eigenstate has wave function
\be
h_n(q)= \frac{1}{\sqrt{\sqrt{\pi}\, 2^{n}\,n! }} 
	\exp(-y^2/2 )H_{n}(y),
\ee
where $H_n(y)$ is a Hermite polynomial of order $n$, and 
$y= (\o/\hb)^{1/2} q$.
An eigenfunction $\psi(\phi)$ of ${\mathcal{H}}_{\mathrm{JJ}}$ with 
eigenvalue $E_J(\nb)$ can be written as 
$\psi(\phi)=\exp(i\,\bar{n}\,  \phi )\,  \eta(\phi)$, 
and satisfies the Schr\"{o}dinger equation,
${\mathcal{H}}_{\mathrm{JJ}}\,\psi(\phi)= E_J \, \psi(\phi)$.
This equation can be written out explicitly using the representation
$n = -i\frac{d}{d\phi}$, which follows from the commutation relation
between $n$ and $\phi$, as
\be
\frac{d^2\,Y(x)}{d\,x^2}+\left(\frac{8\, E_J}{U} +2\, 
Q \cos2\, x \right) Y(x)=0, \label{eq:Mathieu}
\ee 
where $x = \phi/2$, $Y(x)=\eta(\phi /2)$, and $Q= 4\, J/U$. 
This is a Mathieu equation with characteristic value $
a={8\,E_J}/{U}$, and potential of strength $Q$.  The eigenvalues
$E_J$ are determined by the requirement that 
the wave function,  $\psi(\phi)$,  should be single valued, i.\ e., that
$\psi(\phi)=\psi(\phi+2 \pi)$, or equivalently
$
Y(x+\pi)=\exp (- 2\,  i\,  \bar{n}\,  \pi)\,  Y(x).
$
We therefore define $Y_k(x)$ as a Floquet solution of the Mathieu
equation (\ref{eq:Mathieu}), with Floquet exponent $\nu = 2\, k -2\,
\nb$, where $ k=0, \pm 1, \pm 2, \ldots $. We denote the corresponding
eigenvalue of ${\mathcal{H}}_{\mathrm{JJ}}$ by $E_{J,k}(\nb)$.  For $0
< \nb < 0.5 $, the lowest eigenvalue corresponds to $k=0$, followed in
order by $k = 1,-1,2,-2,\ldots $.  The eigenvalues of
${\mathcal{H}}_{\mathrm{JJ}}$ are periodic in $\nb$ with period unity,
and are also symmetric about $\nb=1/2$.

Finally, any eigenstate $\Psi(\phi, q)$, of the Hamiltonian ${\mathcal{H}}$ 
can be expressed as a linear combination of product wave functions consisting
of eigenfunctions of ${\mathcal{H}}_{\mathrm{JJ}}$ and 
${\mathcal{H}}_{\mathrm{photon}}$:
\be
\Psi(\phi, q)= \sum_{k,n} A_{kn} \psi_k(\phi)\,h_n(q) \label{eq:psi},
\ee
where $\psi_k(\phi)= \exp (i\, \nb\, \phi) Y_{k}(\phi/2)$, and $A_{k n}$
are the expansion coefficients.
The only term in ${\mathcal{H}}$ which is not diagonal in this product basis is the
interaction term ${\mathcal{H}}_{\mathrm{int}}$.  The eigenfunctions and eigenvalues of
the Hamiltonian of Eq. (\ref{eq:htot}) follow from 
the Schr\"{o}dinger equation,
$
{\mathcal{H}}\,\Psi(\phi,q)= E\,\Psi(\phi,q).
$

\subsection{Canonical Transformation}

Before proceeding to the solutions of this Schr\"{o}dinger equation,
we show that ${\mathcal{H}}$ is equivalent, via a canonical
transformation, to another Hamiltonian which has sometimes been
discussed in the literature \cite{shnirman,hekking}.  A similar
transformation has been demonstrated in Ref. \cite{shnirman} for the
case of one and two junctions.  The required transformation is as
follows:
\be
\left\{
\begin{array}{ll}
\phi^\prime &=\phi-g \sqrt{\o/\hb} \, q \\
 n^\prime &= n  \\
p^\prime &= p+ g \sqrt{\hb\, \o }\, n  \\
q^\prime &= q.  \\
\end{array}
\right.
\ee
It is readily found that the new operators satisfy the commutation
relations $[\phi^\prime, n^\prime]=i$, $[q^\prime, p^\prime]
=i\hb$, with all other commutators vanishing.
The Hamiltonian ${\mathcal{H}}$ can readily be expressed in 
terms of the new variables.  The result is
\bea
{\mathcal{H}}' =
\half U^\prime (n^\prime&-&\nb^\prime)^2 -J \cos\,\phi^\prime + 
\half( p^{\prime \, 2}+ \o^2 q^{\prime\, 2})\nn \\
&-& g \sqrt{\hb\, \o }\,
p^\prime n^\prime + 
\frac{ U\,\hb\, \o  \,\nb^2 \,g^2}{2\,( U+ \hb\, \o\, g^2 )}\label{eq:hamTC},
\eea
where  
$\nb^\prime =  U \,\nb/(U + \hb \,\o\, g^2)$
and $U^\prime = U + \hb \,\o\, g^2$.

If we interpret the photonic mode as an LC resonator, then this
Hamiltonian describes a junction which is {\em capacitively} coupled
to that mode.  Obviously, this transformed Hamiltonian will have the
same spectrum of eigenvalues as the original one.  Although it is not
obvious from the form of Eq.\ (\ref{eq:hamTC}), we have numerically
confirmed that the spectrum is symmetric with respect to $\bar{n} =
1/2$, as is implied by the original Hamiltonian.

\section{Numerical Results}

We have diagonalized ${\mathcal{H}}$ [Eq.\ (\ref{eq:htot})]
numerically, using the basis discussed in Section IIB.  We used a
truncated basis which is composed of the direct product of the lowest
five energy levels of the junction ($k=0,\pm1,\pm2 $), and the lowest
ten of the electromagnetic field ($n=0,1,\ldots,9 $).  We have also
confirmed numerically that increasing the number of states in this
basis has little effect on the eigenvalues at least up to $E = 6\, U$.

Fig.\ 1 shows the calculated eigenvalues of ${\mathcal{H}}$, plotted
as a function of $\bar{n}$ up to an energy of $U$, for coupling
constants $g = 0$, $g = 0.15$ and $g = 1.5$.  In all three cases, we
have arbitrarily chosen $Q = 0.7$ and $\hb \,\o/U= 0.3$, corresponding
to a case where the charging energy, Josephson energy, and quantum
$\hbar\,\omega$ of radiation energy are all comparable in
magnitude. The degeneracy of some of the energy levels in Fig.\ 1(a)
is broken in Fig.\ 1(b) (the numerical value of the second and third
energy levels splitting at $\nb=0.258$ is given in the caption); the
degeneracy is more noticeably broken in Fig.\ 1(c), where the coupling
constant $g$ is much larger.  The degeneracy-breaking is, of course,
caused by the perturbation ${\mathcal{H}}_{\mathrm{int}}$.  Moreover,
the magnitude of the energy splitting also tends to increase with
increasing $g$.  This behavior is a characteristic feature of ``level
repulsion'' predicted by standard degenerate perturbation theory in
quantum mechanics.

Next, we discuss the time evolution of the system described by Hamiltonian
(\ref{eq:htot}), given its state at time $t = 0$.  The expectation value
of any operator ${\mathcal{O}}$ at time $t$ is
\be
\langle {\mathcal{O}}(t)\rangle = \mathrm{Tr}(\rho(t)\mathcal{O}),
\ee
where $\rho(t)$ is the density matrix of the system at time $t$.  $\rho(t)$ is
given by
$
\rho(t)= U(t) \rho(0) U^{\dagger}(t),   
$
where $U(t)= \exp(-i{\mathcal{H}}t/\hb)$
is the evolution operator, and $\rho(0)$ is determined by the 
initial state of the system.  The time average
of the operator ${\mathcal{O}}$ is then given by
\be
\la \la {\mathcal{O}}\ra\ra= \lim_{T\rightarrow\inf} \frac{1}{T}\int_{0}^{T}
\mathrm{Tr}\,(\rho(t) \,{{\mathcal{O}}}) \, {dt}, \label{eq:timeA}
\ee
where the inner and outer set of triangular brackets denote, respectively,
a quantum mechanical and a time average.

As an illustration, we have calculated 
$\la\la{\mathcal{H}}_{\mathrm{JJ}}\ra\ra$ and  $\la\la{\mathcal{H}}_{\mathrm{photon}}\ra\ra$.
For each operator, we carried out the calculation making the arbitrary
assumption that 
the state of the system at  time $t=0$ is
$|\alpha=i \sqrt{2}\,;\, k \ra$.  
Here  
$|\alpha=i \sqrt{2}\, \ra $ is a coherent state of the electromagnetic field,
i.\ e., an eigenstate of the annihilation operator, $a$ such that 
$a |\alpha\ra= \alpha |\alpha \ra$.  
(We have  confirmed numerically that the time averages are unchanged
if the initial state is an eigenstate of the number operator $a^\dag a$.)
$| k \ra$ is an eigenstate of ${\mathcal{H}}_{\mathrm{JJ}}$ 
with quantum number $k$, and we have considered two
different initial states, corresponding to $k = 0$ and $k = 1$.
The density matrix $\rho(0)$ at time $t=0$ is easily calculated
once the initial state is specified; in obvious notation it is 
$
\rho(0)= |\alpha=i \sqrt{2}\,;\, k \ra 
\la \, \alpha= i \sqrt{2}\,;\, k | . 
$
We have calculated the time evolution using the parameters 
$Q=0.7,\, \hb\, \o /U= 0.3$, and  $g=0.15$. 

The time-averaged energy $\langle\langle
{\mathcal{H}}_{\mathrm{JJ}}\rangle\rangle$ for these parameters is
shown in Fig.\ 2a for $k = 0$ and $k = 1$.  Each curve shows a strong
structure near $\nb = 0.26$ (and $\nb = 0.74$).  At these values of
$\nb$, according to Fig.\ 1, the difference in energy between the
ground state energy, and the first excited energy is $ \sim 0.3 \, U =
\hbar\,\omega$.  The structure thus corresponds, we believe, to a {\em
resonant absorption} of a photon by a Cooper pair in the junction -
that is, at this value of $\nb$, Cooper pairs of electrons move from
the ground state, $|k=0\ra$, of the junction to the first excited
state, $|k=1\ra$.  The corresponding structure for $k=1$ (dotted line)
arises from stimulated emission, in which a Cooper pair falls from the
excited state to the ground state of the junction with the emission of
a photon.  The exchange of energy between the junction and the
resonator can also be seen in Fig.\ 2b, where we show the
time-averaged energy contained in the photon mode, $\la\la
{\mathcal{H}}_{\mathrm{photon}} \ra\ra $, for the same two initial
states. $\la \la {\mathcal{H}}_{\mathrm{photon}} \ra \ra$ for both
states is close to the unperturbed value $(\la n \ra +1/2) \hb\, \o =
5 \,\hb \,\o/2$, which is the energy contained in the photon field for
the coherent state $|\alpha= i \sqrt{2} \ra $, but one increases
sharply, while the other decreases, near the resonance at $\nb=0.26$
(and $\nb=0.74$).

The curves in Fig.\ 2 also show weaker structure near $\nb \approx
0.06$ ( and near $0.94$ ). This structure corresponds to a process
in which {\em two photons} are absorbed by the junction Cooper pairs.
Specifically, the junction is excited from its ground state to its
second excited state at this value of $\nb$, with a consequent loss of
two photons from the electromagnetic field. This is discussed further
below.

We have also studied the time evolution of this system under
conditions such that $\nb$ is fixed at a value of $0.258$, near the
principal resonance mentioned above.  To see the time evolution, we
prepared the system at time $t=0$ in the state $|n=1 \,;\, k=0 \ra$,
for which the photon resonator is in state $n = 1$ and the junction in
state $k = 0$.  We then allowed the system to evolve in time according
to the evolution operator $U(t)=\exp(-i{\mathcal{H}}\,t/\hb)$.  In
Fig. 3 we show the resulting time-dependent expectation value of $\la
{\mathcal{H}}_{\mathrm{JJ}}\ra$, and of $\la
{\mathcal{H}}_{\mathrm{photon}}\ra$.  Also shown in the same figure
are the time-dependent probabilities for finding the junction in state
$k = 1$ at time $t$ and the resonator in state $n = 1$ at time t,
given this initial state.  Clearly, the system is oscillating between
the states $|n=1\,;\, k =0 \ra$ and $|n=0 \,;\, k =1 \ra$.  This
energy exchange is periodic in time: both $\la
{\mathcal{H}}_{\mathrm{JJ}}\ra$ and $\la
{\mathcal{H}}_{\mathrm{photon}}\ra$ vary periodically in time, with a
period $T_1 \approx 675 \,\hb /U $ for our particular choice of $Q/U$
and $\hb \, \o /U$.  In the first half of this period, Cooper pairs
absorb energy from the electromagnetic field and are driven into the
first excited state, while in the second half, they lose energy to the
electromagnetic field and fall back into the ground state.  In the
present model, this process continues indefinitely.  In a more
realistic model which includes dissipation, these oscillations would
gradually decay in time.

Note also that the existence of a periodic oscillation between one
state and another is not dependent on the choice of the initial state.
For example, if the initial state is chosen as $|n=0\,;\ k =1 \ra$,
the junction would interact with the vacuum fluctuations of the
electromagnetic field, and would periodically visit the state
$|n=1\,;\, k =0 \ra$.  We have confirmed this behavior numerically in
our model.

In Fig.\ 4, we show $\langle {\mathcal{H}}_{\mathrm{JJ}} \rangle$ and
$\langle {\mathcal{H}}_{\mathrm{photon}} \rangle$ as a function of
time for two different initial states: $|n=1 \,;\, k =0 \ra$ (full
line) and $|n=1 \,;\, k =1 \ra$ (dashed line).  (The first of these is
also shown in Fig.\ 3.)  Evidently both states oscillate periodically
in time, but with slightly different periods.

Fig.\ 5 \,shows the analogous time evolution of a system prepared in a
fixed initial state at $\nb = 0.061$, where we expect to see a
signature of a two-photon absorption.  In this case, we prepared the
system in the initial state $|n=2\,;\, k =0 \ra$.  The system
oscillates between the $k = 0$ (ground state) and $k = -1$ (second
excited state) of the Josephson junction, which is separated from the
ground state by an energy $2\,\hbar\,\omega$ at this value of $\nb$.
The number of photons in the system also oscillates periodically in
time, between the values $n = 2$ and $n = 0$, with period $T_2 \approx
7500 \, \hb /U $.  This period is acd bout eleven times longer than the
period at $\bar{n} = 0.258$ (single-photon absorption), because the
matrix element connecting the two states at $\bar{n} = 0.06$ is much
smaller.

\section{Discussion}

The problem studied here closely resembles that of a two-level atom
interacting quantum-mechanically with a monochromatic electromagnetic
field.  Specifically, when $\nb$ is such that the field is
approximately in resonance with a transition between the lowest and
first excited state of the Josephson junction, then it is a good
approximation to consider only four states in the basis.  We now show
that this approximation gives results in good agreement with those
obtained using the full direct product basis.

We consider a basis constructed out of the two lowest energy states of
the junction, i.\ e., for $ k=0\, , \, 1$ and of the electromagnetic
field, for $n=0\, , \,1 $.  Using a four dimensional basis formed from
the direct product of these states, $|n=0,1\,;\, k=0, 1 \ra$, we can
express the total Hamiltonian of Eq. (\ref{eq:htot}) as a four times
four dimensional matrix.  If we retain in the basis only the two
states $|0\,;\, 1\rangle$ and $|1\,;\, 0\rangle$ which are
approximately degenerate, the Hamiltonian can be further reduced to
the $2 \times 2$ matrix
\bea
{\mathcal{H}} = 
	\epsilon_{0,1} \,| 0\,;\,1 \ra\la &0&;\,1| +
		\epsilon_{1,0}\, | 1\,;\,0 \ra \la 1\,;\,0|\nn \\ 
	&+& \xi_1\,\, | 0\,;\,1 \ra \la 1\,;\,0| 
		+ {\xi}^{*}_{1} | 1\,;\,0 \ra \la 0\,;\,1|.  
\label{eq:hamReduced}
\eea

The diagonal matrix elements take the values $\epsilon_{0,1}=
E_{J,1}(0.258)+ \hb \,\o/2 $; $\epsilon_{1,0}= E_{J,0}(0.258)+ 3\, \hb
\,\o/2 $, where $E_{J,k}(\nb)$ is the k$^{th}$ eigenenergy of
${\mathcal{H}}_{\mathrm{JJ}}$ at $\nb$.  $\xi_1$ is given by $\xi_1=
\la 0,1 |\frac{1}{2}{\mathcal{H}}_{\mathrm{int}}|1,0 \ra $, where
${\mathcal{H}}_{\mathrm{int}}$ is given by Eq.\ (\ref{eq:hint}).  To
first order in the coupling strength $g$,
${\mathcal{H}}_{\mathrm{int}} \approx - J \, g \sqrt{(\o/\hb)}\, q
\sin \phi $, and hence, to the same order in $g$,
$
\xi_1 \approx -({J\,g}/{\sqrt{8}}) \, \la k=0| \sin\phi |k=1 \ra .
$
A similar two-level approximation can be carried out for coupling between
the states $|n;\,1\rangle$ and $|n-1;\, 0\rangle$; 
in this case, the coupling energy is found to be $\xi_1\,\sqrt{n}$.  However,
as expected, we have found that the two-level approximation becomes 
progressively more inaccurate as $n$ is increased.

To illustrate this approximation, we consider the case where the
energy levels $\epsilon_{0,1} $ and $\epsilon_{1,0}$ are degenerate.
Diagonalizing the reduced Hamiltonian matrix, Eq.\
(\ref{eq:hamReduced}), gives us the two energy levels $E_{\pm}=
\epsilon_0 \pm |\xi_1|$.  These levels correspond to two entangled
photon-Josephson states.  The Rabi period of oscillation between them
is $T_1 = \pi \hb / |\xi_1| $.  Using the parameters and wave
functions corresponding to our calculation of Fig.\ 3, we obtain $T_1
\approx 667 \, \hb /U$, which is in excellent agreement with the
value found from the full numerical solution ($675 \, \hb /U$), based
on the complete basis and shown in Fig.\ 3.  We conclude that indeed
the reduced Hamiltonian, defined in Eq.\ (\ref{eq:hamReduced}), is an
excellent description of the system, at least when the two lowest
levels are close to resonance.

The two-photon process discussed in Fig.\,6  can also  
be understood  using the
reduced Hamiltonian:
\bea
{\mathcal{H}} = \epsilon_{2,0} \,| 2\,; &0& \ra\la 2;\,0| 
	     + \epsilon_{0,-1}\, | 0\,;\,-1 \ra
		\la 0\,;\,-1|\nn \\ 
	     &+& 
		\xi_2\,\, | 2\,;\,0 \ra \la 0\,;\,-1| 
		+ \xi_{2}^{*} | 0\,;\,-1 \ra \la 2\,;\,0|.  
\label{eq:hamReduced2}
\eea
Here we consider only the two degenerate levels with energy,
$\epsilon_{2,0}=E_{J,0}(0.061)+ 5 \,\hb \,\o/2 $;
$\epsilon_{0,-1}=E_{J,-1}(0.061)+ \hb\, \o/2 $. The coupling strength
$\xi_2$ is given by $\xi_2= \la 2,0
|\frac{1}{2}{\mathcal{H}}_{\mathrm{int}}|0,-1 \ra $, where
${\mathcal{H}}_{\mathrm{int}}$ is given by Eq.\ (\ref{eq:hint}).  To
second order in the coupling strength $g$,
${\mathcal{H}}_{\mathrm{int}} \approx - (1/2) J \, g^2 (\o/\hb)\, q^2
\cos \phi $, and hence, to the same order in $g$, $ \xi_2 \approx
-({J\,g^2\,\sqrt{2}}/{8}) \, \la k=-1| \cos\phi |k=0 \ra .  $ Using
the same parameters as before, we find that the Rabi oscillation
period is $T_2= \pi \hb /|\xi_2| \approx 5800 \,\hb/U $, compared to
the value $7500\, \hb/U$ obtained from the exact numerical solution.
Thus the two-level approximation is not quite as  good as in the
one-photon case, at least for these parameters.

As mentioned in the Introduction, our \h model shows the phenomenon of
``collapse and revival,'' provided that the resonator is prepared
initially in a coherent state\cite{eberly}. This behavior arises from
the interference between the different Rabi oscillations which exist
when the initial state is a superposition of different photon number
states. In Fig. 6. we show the time dependent probability that the
junction is in the excited state, $k=1$, when the initial state of the
system is $| \alpha=i \sqrt{3} ; k=0 \ra$. Also on the same figure we
show (dotted line ) the same probability calculated using the reduced
\h of Eq. (\ref{eq:hamReduced})\cite{eberly}:
\be
{\mathcal{P}}(t)=\half-\half\, e^{-|\alpha|^2}\,\sum_{s=0}^{\infty} 
\frac{{|\alpha|}^{2\,s}}{s!} \cos( 2\, \xi_1\,\sqrt{s} \,t).\label{eq:CoRev}
\ee
The agreement between the two results is very good, especially for short
times. For longer times or for larger $\alpha$, the agreement
between the two level approximation and the full numerical solution is
less good, primarily because the number of photons in the cavity
becomes large.

Before concluding, we make some order-of-magnitude estimates for some
of the parameters entering our model, and compare them with those in
plausible junctions and cavities. First, the parameter $g$ may be
estimated using Eq.\ (\ref{eq:g}). If we arbitrarily assume a
cylindrical cavity of radius $r$ and length $d$, the frequency of the
lowest mode is approximately $2\,c/r$, and may be either a TE or TM
mode, depending on the ratio $d/r$. Substituting into Eq.\
(\ref{eq:g}), we obtain $g \sim \ell_\| \sqrt{8/(137\,r\,d)} \sim
0.24\,\ell_\|/\sqrt{r\,d}$, where $e^2/(\hbar c) \approx 1/137$ is the
fine structure constant. 
To estimate $\hbar\,\omega/U$, we use the same expression for $\omega
\approx 2\,c/r $ and $U= 4\,e^2/C$, where $C$ is the junction
capacitance.  Taking $C = x$, where $x$ is a characteristic length, we
obtain $\hbar\,\omega/U \sim 137\,x/r$.  Thus, a value of $0.3$ for
this parameter implies $x/r \sim 0.002$ or less.  For a capacitance of
$10^{-16}$~F, readily achievable within current
technology\cite{shnirman}, we have $x \sim 10^{-3}$~cm, or $U \sim 1
\times 10^{-15}$~erg~$\sim 8$~K, and hence $r \sim 0.5 $~cm. The
associated frequency would be $\omega \sim 100$~GHz, which is in the
range used in some recent experiments\cite{barbara}. Of course the
quantum of radiation is limited by  the superconducting gap
$\Delta$ of the superconductor used. We also note that our choice $Q =
0.7$ corresponds to $J \sim 3$~K for the above parameters, or
equivalently, a critical current $I_c$ of about $0.1~\mu$A. The value
of $J$ should be of the same order as a quantum of radiation, $\hb \,\o$.

Using the above estimates, and assuming a realistic junction with an
insulating layer of thickness $\ell \sim 100 $~nm, the coupling
strength $g$ will be $g \sim 10^{-5}$ if $ r \sim d \sim 0.5$~cm. If
we make $d \ll r$, as in a disk-like microcavity, we can presumably
increase $g$ substantially, but it will still remain much below the
value of $0.15$ used in our calculations. Such a value of $g=0.15$, or
even larger, is preferable because this would make the Rabi period of
the same order or less  than the coherence time in Josephson junctions. In a
high-Q cavity, the dissipation-induced decoherence time may be as
large as $2$~ns\cite{nakamura2}, much smaller than would be produced
by $g \sim 10^{-5}$. A larger $g$ would decrease the Rabi time, and
technological improvements will no doubt increase the junction
coherence time, making these times more comparable. Another way to
increase the effective coupling, and hence decrease the Rabi period,
would be to use many identical junctions located in a region small
compared to a wavelength. The effective coupling is then found to
increase as the square root of the number of junctions in the cavity
\cite{alsaidi}.  On the basis of all these estimates, we conclude that
the two-level approximation tested in this paper should be excellent
for the range of parameters likely to be achievable experimentally.
Of course, the precise relevant parameters for a given experiment
depend on the details of the geometry, which have only been very
crudely estimated in the above discussion.
 
It is of interest to compare the present quantum problem to another
system which bears some similarities, namely an {\em array} of
Josephson junctions placed in a resonant cavity\cite{barbara}.  This
system involves large numbers of junctions, each coupled to a
single-mode cavity such as is considered here.  While the appropriate
Hamiltonian for treating such an array resembles that of the present
problem, there are important differences.  First, the observed array
behavior involves large numbers of photons in the microcavity, whereas
the present model is suitable to the case of one or a few
photons. Secondly, the IV characteristics of the array involve
junctions which are biased onto either the resistive or the
superconducting branch, whereas the present model considers only the
latter branch. Finally, on the basis of the above estimates, the
junctions to which the present calculation applies are probably
substantially smaller, with smaller critical currents, than those
studied in Ref. \cite{barbara}. Thus, a rather different approach,
such as those described in Refs.\
\cite{filatrella,cawthorne,almaas,bonifacio,harbaugh} is called for.

To summarize, we have considered the interaction between a small
Josephson junction and the photon field in a resonant cavity, in the
limit when dissipation can be neglected.  We find that there is a
strong interaction between the junction and the resonant photon mode
at special values of the gate voltage, as defined by the variable
$\bar{n}$.  At such values of $\bar{n}$, the lowest two eigenstates of
the system correspond to {\em entangled states} involving states of
both the photon field and the Josephson junction.  The Rabi period of
oscillation between these states is inversely proportional to a
certain coupling strength $g$.  We also found evidence for a
two-photon absorption process at another resonant value of the
variable $\bar{n}$.  Finally, we found that the lowest entangled
eigenstates are accurately described by a simple two-level model which
is a truncation of the full model Hamiltonian, Eq.\ (\ref{eq:htot}).

We conclude with a brief discussion of possible improvements in the
present calculations.  First, of course, the effects of dissipation
need to be included.  In principle such dissipation can be included by
coupling the Hamiltonian degrees of freedom to a bath of harmonic
oscillators.  If the density of states of this bath has the right
frequency-dependence, this coupling leads to Ohmic
damping\cite{caldeira,ambeg,chakra}.  Secondly, it would be of great
interest to extend this work to more than one junction coupled to the
same resonant cavity\cite{alsaidi}. If the two-level
approximation continues to be accurate in this case, the study of such
a group of  Josephson junctions would be greatly simplified.

\section{Acknowledgments}  We are grateful for useful discussions with
E. Almaas. This work has been supported by the National Science
Foundation, through Grants DMR97-31511 and DMR01-04987.


\begin{center}
FIGURE CAPTIONS
\end{center}

\vspace{0.1in}

\begin{enumerate}

\item Lowest eigenvalues $E/U$ of the system described by the
Hamiltonian (\ref{eq:hint}), calculated as a function of $\bar{n}$ for
the parameters $\hb\, \o /U =0.3$ and $Q = 0.7$, and (a) $g = 0$, (b)
$g = 0.15$, and (c) $g = 1.5$. All eigenvalues are shown up to $E/U =
1$. Note in particular the breaking of the degeneracy of the second
and third energy levels around $\nb=0.26$, seen clearly in
Fig. 1c. The separation $\Delta E$ between these 
levels at $\nb=0.258$ is $\Delta E/ U = 0,\, 0.01,$ and $0.06$ in
(a), (b) and (c) respectively.

\item 
Time and quantum-mechanical average of the operator 
${\mathcal{H}}_{\mathrm{JJ}}$ [Fig. 2(a)], and 
of ${\mathcal{H}}_{\mathrm{photon}}$ 
[Fig.\ 2(b)], plotted as a function of $\bar{n}$ for two different
assumptions about the initial state of the system:
$|\alpha=i \sqrt{2}\, ;\, k=0 \ra$ (solid curves); and
$|\alpha=i \sqrt{2}\,;\,k=1 \ra$ (dashed curves). The states
are defined explicitly in the text.
In both cases, we used the parameter values
$\hb \,\o /U =0.3$, $Q = 0.7$, and $g=0.15 $.

\item (a) Time-dependent value of 
$\langle {\mathcal{H}}_{\mathrm{JJ}} \rangle$ (full curve) and
of $\langle {\mathcal{H}}_{\mathrm{photon}}\rangle $ (dashed curve) 
for $\nb = 0.258$. The system is prepared at time $t = 0$ in state $| n=1 \,;\,k=0 \ra $.  
The other parameters are $Q = 0.7$, $\hbar\omega/U = 0.3$, and $g = 0.15$.  
(b) Same as (a) except that we show the time-dependent
probability for finding the Josephson junction (full curve) and
the photonic resonator (dashed curve) in their first excited
states.

\item (a) Time-dependent value of $\langle {\mathcal{H}}_{\mathrm{JJ}}\rangle$
for two different initial states: $| n=1 \,;\, k=0 \ra $ (full curve), 
and  $| n=1 \,;\, k=1 \ra $ (dashed curve).  In both cases, $\nb = 0.258$,
$Q = 0.7$, $\hbar\,\omega/U = 0.3$, and $g = 0.15$.  (b). Same as (a) except
we plot $\la {\mathcal{H}}_{\mathrm{photon}} \ra$.

\item Same as Fig. 3, except that $\nb$ is fixed at $0.061$, and the
system is prepared initially in the state $| n=2\,;\, k=0 \ra $ at
time $t=0$.  Other parameters are the same as in Fig.\ 3.

\item Time dependent probability that the junction is
in the first excited state  given that initially  
the resonator is in the coherent state $|\alpha= i
\sqrt{3} \ra$ and  the junction  in the ground state. 
Also shown (dotted line) is the probability ${\mathcal{P}}(t)$
[Eq. (\ref{eq:CoRev})] calculated using the 
 two level approximation. Here we used $\nb = 0.258$,
$Q = 0.7$, and $\hbar\,\omega/U = 0.3$.
 
\end{enumerate}



\begin{figure}[tb]
\epsfysize=3cm
\centerline{ \epsffile{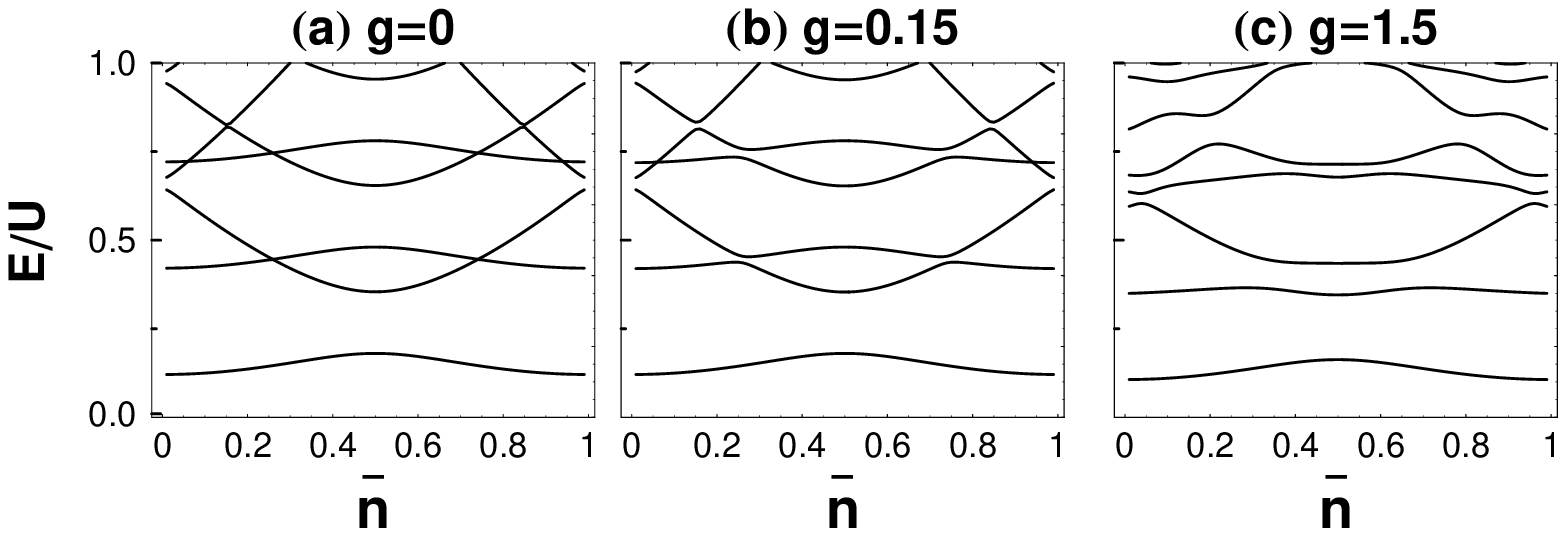}}
\label{fig1}
\end{figure}

\begin{figure}[tb] 
\epsfysize=3cm
\centerline{ \epsffile{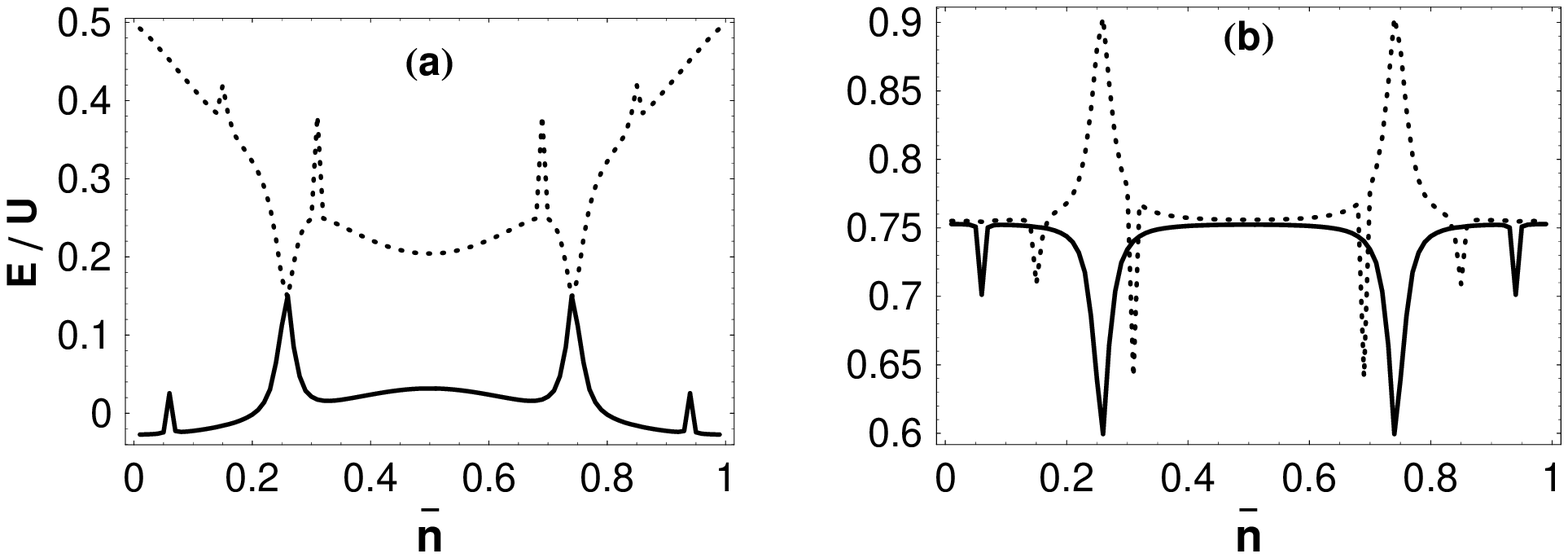}}
\label{fig2}
\end{figure}

\begin{figure}[tb] 
\epsfysize=3cm
\centerline{\epsffile{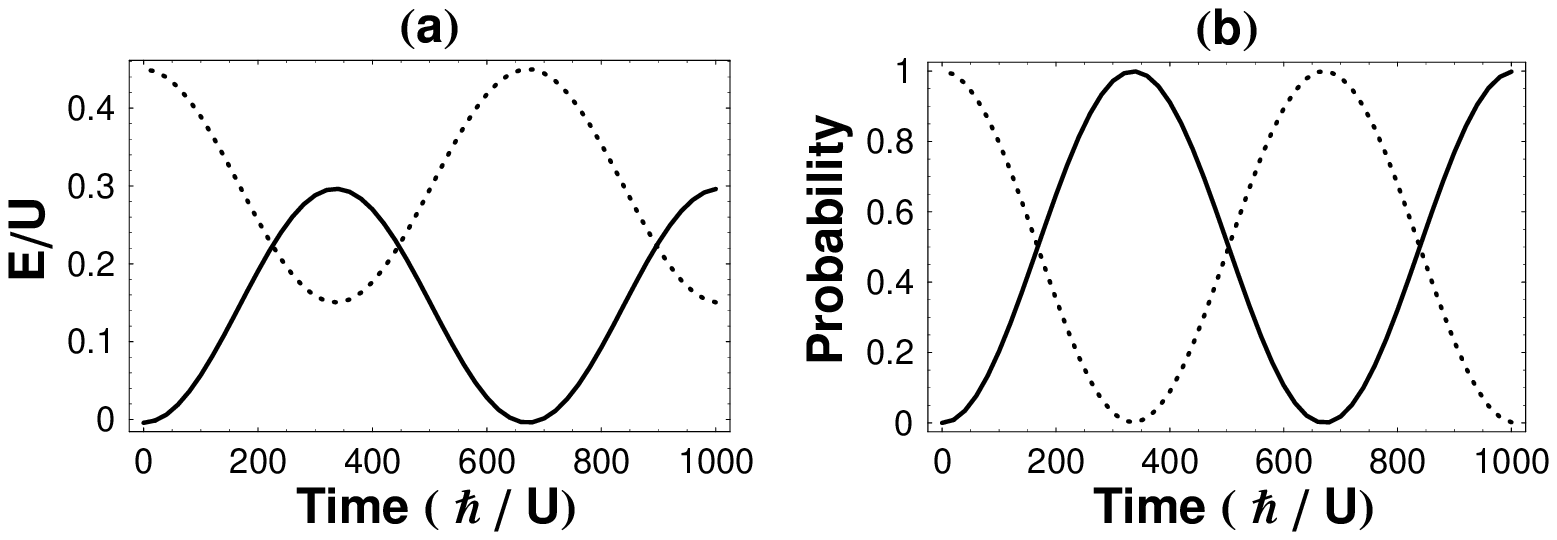}}
\label{fig3}
\end{figure}

\begin{figure}[tb] 
\epsfysize=3cm
\centerline{ \epsffile{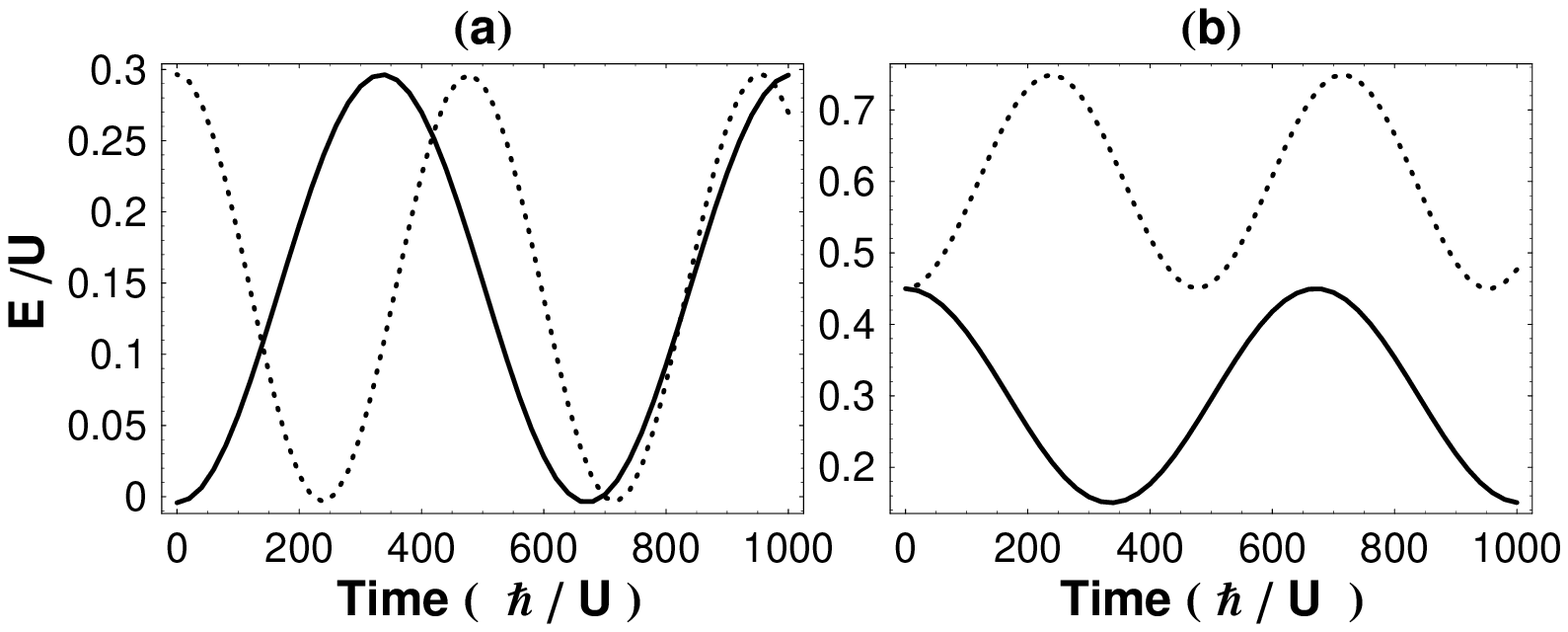}}
\label{fig4}
\end{figure}

\begin{figure}[tb] 
\epsfysize=3cm
\centerline{ \epsffile{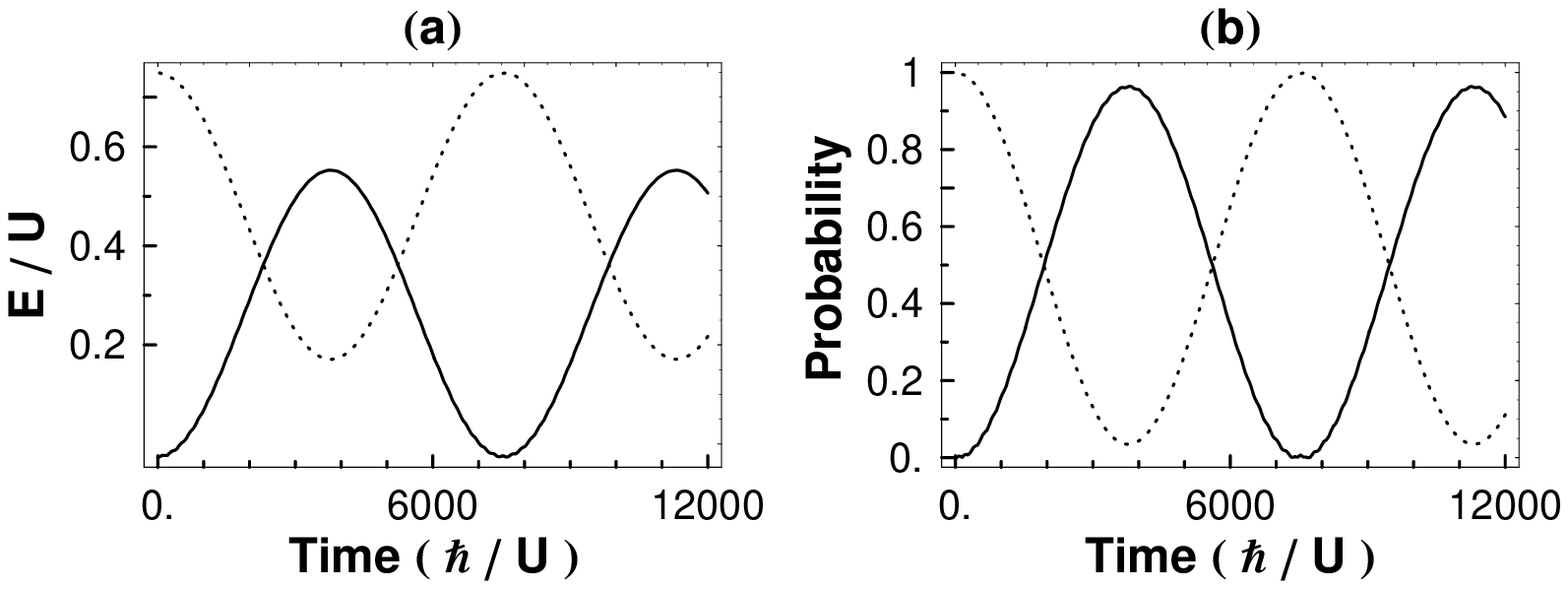}}
\label{fig5}
\end{figure}

\begin{figure}[tb] 
\epsfysize=4cm
\centerline{ \epsffile{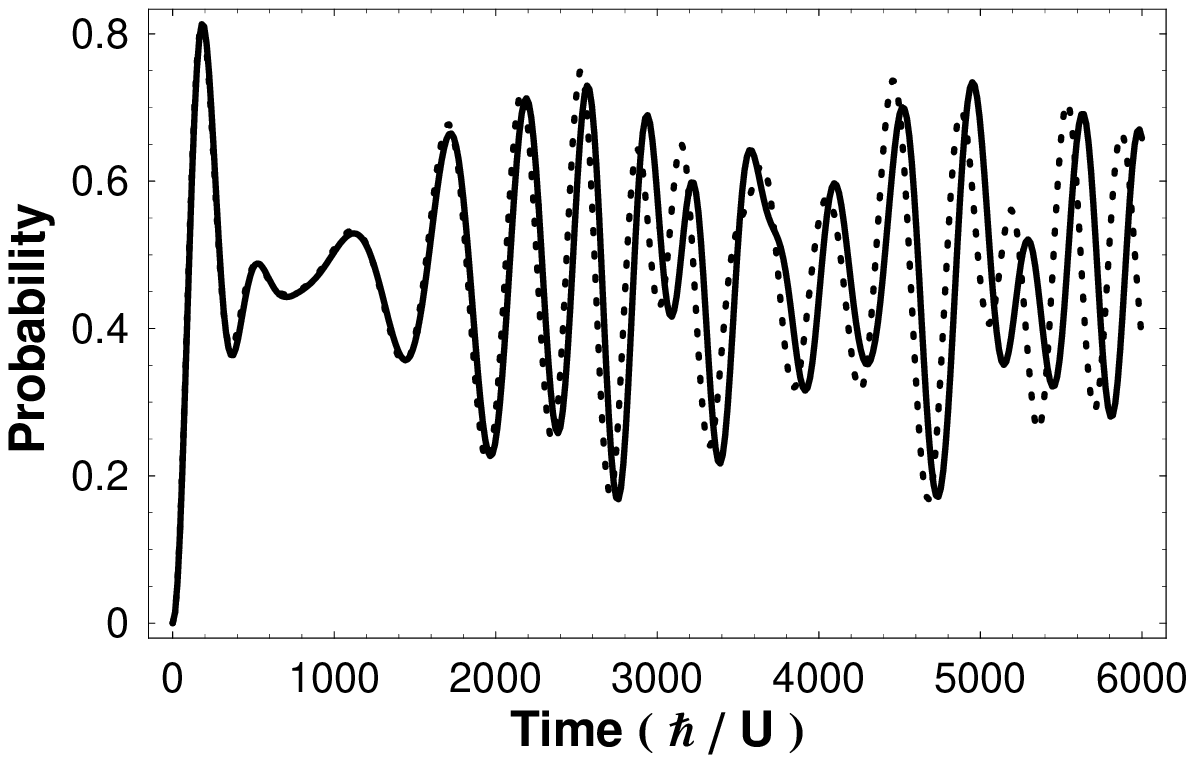}}
\label{fig6}
\end{figure}


\begin{thebibliography}{3}
\bibitem[*]{email1} Al-Saidi.1@osu.edu.

\bibitem[\dag]{email2} Stroud@mps.ohio-state.edu.
\\

\bibitem{friedman} J. R. Friedman, V. Patel, W. Chen, S. K. Tolpygo,
and J. E. Lukens, Nature {\bf 406}, 43 (2000).

\bibitem{vanderwal} C. H. van der Wal, A. C. J. ter Haar, F. K. Wilhelm,
R. N. Schouten, C. J. P. M. Harmans, T. P. Orlando, S. Lloyd, and
J. E. Mooij, Science {\bf 290}, 773 (2000).

\bibitem{bouchiat} V. Bouchiat, D. Vion, P. Joyez, D. Esteve,
M. H. Devoret, Physica Scripta T{\bf 76}, 165 (1998).

\bibitem{nakamura1} Y. Nakamura, C. D. Chen, and J. S. Tsai, Phys. Rev. Lett. 
{\bf 79}, 2328 (1997). 


\bibitem{nakamura2} Y. Nakamura, Y. A. Pashkin, and J. S. Tsai, Nature {\bf
398}, 786 (1999)

\bibitem{zorin} A. B. Zorin, S. V. Lotkhov, S. A. Bogoslovsky, and
J. Niemeyer, cond-mat/0105211.

\bibitem{mooij} J.\ E.\ Mooij, T. P. Orlando, L. Levitov, L. Tian,
C. H. van der Wal, and S. Llyod, Science {\bf 285}, 1036 (1999).

\bibitem{feigelman} M. V. Feigel'man. L. B. Ioffe, V. B. Geshkenbein,
and G. Blatter, J. Low Temp. Phys.  {\bf 118}, 805 (2000).

\bibitem{steane} For a recent review, see, e.\ g., A.\ Steane,
Rep.\ Prog.\ Phys.\ {\bf 61}, 117 (1998).

\bibitem{haroche} S. Haroche, Physics Today {\bf 36}, 51 (1998).

\bibitem{shnirman} A. Shnirman, G. Sch\"{o}n, and Z. Hermon, 
Phys. Rev. Lett. {\bf 79}, 2371 (1997).

\bibitem{hekking} O. Buisson and F. W. J. Hekking, cond-mat/0008275
(2000). 

\bibitem{everitt} M. J. Everitt, P. Stiffell, T. D. Clark, A. Vourdas,
J. F. Ralph, H. Prance, and R. J. Prance, Phys. Rev. B {\bf 63},  144530 (2001).

\bibitem{eberly} J. H. Eberly, N. B. Narozhny, and J. J. Sanchez-Mondragon, Phys. Rev. Lett. {\bf 44}, 1323 (1980).

\bibitem{barbara} P. Barbara, A. B. Cawthorne, S. V. Shitov, and C. J.
Lobb, Phys. Rev. Lett. {\bf 82}, 1963 (1999). 
 
\bibitem{alsaidi} W.~A.~Al-Saidi, and D.~ Stroud, In preparation.

\bibitem{filatrella} G. Filatrella, N. F. Pedersen, and K. Wiesenfeld,
Appl. Phys. Lett. {\bf 72}, 1107 (1998); and Phys. Rev. E{\bf 61},
2513 (2000).

\bibitem{cawthorne} A. B. Cawthorne, P. Barbara, S. V. Shitov,
C. J. Lobb, K. Wiesenfeld, and A. Zangwill, Phys. Rev. B{\bf 60}, 7575
(1999).

\bibitem{almaas} E. Almaas and D. Stroud, Phys. Rev. B{\bf 63},
144522 (2001), and submitted to Phys. Rev. B.  Note that the coupling
constant ``$g$'' as defined in this reference is proportional to the
square of the $g$ defined here.

\bibitem{bonifacio} R. Bonifacio, F. Casagrande, and M. Milani, Lett. al
Nuovo Cimento {\bf 34}, 520 (1982); L. A. Lugiato and M. Milani, Nuovo
Cimento B{\bf 55}, 417 (1980).

\bibitem{harbaugh} J. K. Harbaugh and D. Stroud, Phys. Rev. B{\bf 61},
14765 (2000). 

\bibitem{caldeira} A. O. Caldeira and A. J. Leggett, Ann. Phys. (N. Y.)
{\bf 149}, 374 (1983).

\bibitem{ambeg} V. Ambegaokar, U. Eckern, and G. Sch\"{o}n, Phys. Rev. Lett.
{\bf 48}, 1745 (1982).

\bibitem{chakra} S. Chakravarty, G.-L. Ingold, S. Kivelson, and A. Luther,
Phys. Rev. Lett. {\bf 56}, 2303 (1986). 


\end{thebibliography}
\end{document}